# Effect of Plasmonic Coupling in Different Assembly of Gold Nanorods Studied by FDTD

Aditya K. Sahu* and Satyabrata Raj

*Department of Physical Sciences, Indian Institute of Science Education and Research Kolkata, Mohanpur, Nadia, 741246, India*

*Corresponding author

 E-mail: aks16rs023@iiserkol.ac.in

## Abstract

The influence of the orientation of gold nanorods in different assemblies has been investigated using the Finite Difference Time Domain (FDTD) simulation method. To understand the relative orientation, we vary the size and angle in dimer geometries. Significant effects of plasmon coupling emerged in longitudinal resonances having end-to-end configurations of gold nanorods. The effect of orientational plasmon coupling in dimers gives rise to both bonding and anti-bonding plasmon modes. Effects of various geometries like primary monomer, dimer, trimer, and tetramer structures have been explored and compared with their higher nanorod ensembles. The asymmetric spectral response in a 4 × 4 gold nanorods array indicates a Fano-like resonance. The variation of gap distance in ordered arrays allowed modulation of the Fano resonance mode. The plasmon modes' resonance wavelength and field enhancement have been tuned by varying the gap distance, angular orientation, size irregularity between the nanorods, and nanorod numbers in an array. The integrated nanostructures studied here are not only significant for fundamental research but also applications in plasmon-based devices.

## 1. Introduction

Metallic plasmonic nanostructures have received significant interest for even more than a decade due to their use in enhancement and imaging applications [1-4], nonlinear and quantum optics [5-8], ultra-fast switches [9], and photovoltaic systems [10, 11]. In metallic plasmonic material research, gold nanostructure occupies a prominent place due to its extensive optical properties. Therefore, various works have been done to develop and produce multiple metallic nanostructures to satisfy various application requirements. For all the applications, substantial field enhancement and tuning of surface plasmon resonance are mainly required. It is noteworthy to mention that the most exciting feature of the gold nanorods (GNRs) is their longitudinal surface plasmon resonance (LSPR) [12]. The intensity, frequency, and consistency of the LSPR in an individual GNR is sensitively dependent on a dimension like size and geometry and the surrounding environment (refractive index) along with the presence of other gold nanoparticles nearby [13,14]. Plasmonic structures containing dimers, trimers, and various array geometries have recently received much interest, as they give a tunability of optical properties of the system. The electromagnetic field and the wavelength can be substantially tuned by the separation gap and geometry orientation of the individual components in the plasmonic structure [15-17]. This remarkable ability allows maximum electrical field enhancements around the nanorods by providing a spectrum of light-matter interactions with new mechanisms.

In general, the chemical synthesis method produces a cluster of nanoparticles with much smaller separations with a proper plasmonic response, but the particles inside the cluster can have a wide distribution of different shapes and sizes [18]. Consequently, the assembly of nanoparticles would inevitably involve local irregularities in neighboring particles' geometry and spacing. The LSPR of the combined cluster, particularly in overlapping particles, will be affected in the assembly, enhancing the excitation of charge transfer plasmons [19, 20]. The several new phenomena in dimers and larger structures composed of anisotropic particles make the relative orientation between nanorods an important parameter.

The surface plasmon coupling is considered the hybridization of different surface plasmon resonance modes [21, 22], and the synthesis of ideal dimers in a laboratory is quite difficult. The relative orientation of the nanoparticles in non-ideal dimers produces different gaps in surface morphology. The plasmon coupling induced by organized gold nanorods leads to substantial electric field enhancements and different collective plasmon responses [22]. Studying the optical properties of these structures will help to understand how light interacts with matter. As a consequence, plasmon couplings between metal nanorods can be used to enhance Raman signals, second-harmonics, fluorescence, photoluminescence, and nanometric optical tweezers [23-27]. Metallic nanorods with a close gap can act as optical nanoantennas due to the plasmon coupling-induced large electric field enhancement [28, 29].

In general, surface plasmon absorption bands evolve from the oscillations of conduction electrons in transverse and longitudinal directions. The overall spectra do not give much information about transverse oscillations. Based on its longitudinal surface plasmon resonance, the correlation of a single nanorod into a more extensive assembly is typically studied. Hybridization of plasmon modes can be in-phase or out-of-phase when two metal nanorods interact between themselves. Depending upon the polarization of source light, bonding and anti-bonding modes arise. It has been found that the in-phase hybridization mode, i.e., bonding mode, which is red-shifted in LSPR frequency and strengthens the electric field at the junction of nanorods, arises when the polarized source light is along the nanorod axis. The out-of-phase hybridization mode, i.e., anti-bonding mode, contributes to the blue-shifted LSPR frequency and appears at the non-junction end of the nanorods [14, 30]. However, the LSPR change and plasmon coupling for transverse polarization where the polarization of light is perpendicular to the nanorod axis is minimal [14]. In the ordered array of nanoparticles, the presence of red-shifted narrow peaks is due to the coupling of localized plasmon resonance of individual particles and the collective photonic modes of the ordered array ensembles and can be interpreted as a Fano-like interference model [31]. In some cases, coupling effects generate Fano-like resonances in plasmonic structures. The impact of plasmonic Fano resonances has many applications in meta-materials, signal enhancement, and sensors. [32, 33]. Many simulation studies demonstrate that one can tune the Fano resonance of a structure by varying the incident light polarization [32] and refractive index of the surrounding medium [34], which makes it a good candidate for potential application in chemical sensors [33]. In spite of a lot of studies on Fano resonance in various systems, we have little knowledge on how the morphological change and the separation distances between the nanoparticles in the ordered array are affected by the coupling strength of different plasmonic modes.

The significance of the nanorod orientation on the plasmon coupling has a significant impact; therefore, the effect on the plasmon coupling of the angle between nanorods in dimers is explored. We have investigated the irregularities in coupled LSPRs of nanorod dimers and many different structures using the finite difference time domain (FDTD) simulation method. The theoretical calculations for gold nanorod dimers were carried out on different orientations such as end-to-end, side-by-side, and various array positions to observe their effect with these parameters. In a nanorod assembly, the angular degrees of freedom complicates the prediction of LSPR interactions. Our present studies have simulated the plasmon coupling in Au nanorod dimers by changing the relative angle between two nanorods. We found bonding and antibonding modes due to plasmon hybridization of individual nanorod when one nanorod is rotated with respect to another one in a dimer (with same size nanorods and gap of 3 nm separation). Size mismatch of nanorods in dimers also has a tremendous effect on the optical properties, which is also studied here. A thorough understanding of the effect of size mismatch on the collective optical properties of nanorod dimers has been reviewed here. Theoretically, we also describe a structural geometry of gold nanorods of 4 × 4 matrices, illuminated under normal incidence, presenting a Fano-resonance when separated by a finite gap. The near field aspects, particularly electric field enhancements, have been explored and interpreted with the existing results. We address mainly the plasmon coupling with reliance on four variables: separation, orientation, size heterogeneity, and the number of nanorods in an array of geometry. Correlating the intensities and LSPRs positions in these structural variations of gold nanorod geometries is vital to understand the effect of irregularities in geometry on plasmon coupling, which is an actual scenario in real chemically grown geometry nanorods. To optimize the design and applications of near-field, understanding the plasmon coupling is required and is of prime interest.

## 2. Simulation Methodology

Electromagnetic simulations have been carried out to understand the surface plasmonic properties in various nanorod assemblies by using the FDTD simulation method. FDTD simulation is one of the most common computing tools for numerically solving Maxwell electromagnetic time-dependent equations, and the same has been used to model our nanostructures optically [35]. A preliminary assessment was used to determine the mesh sizes in our simulations to ensure that the numerical results converged. We have used a mesh of 0.5 nm in our simulation for the convergence criteria. We used perfectly matched layer (PML) boundary conditions to prevent any non-physical scattering at boundaries. We used the experimental results of Johnson and Christy to model the complex permittivity of gold [36]. We used the Total Field Scattered Field (TFSF) as the source. We used a power monitor to compute the absorption coefficients inside the TFSF source and another for the scattering coefficients outside the source. The polarization of incident light has been set along the nanorods to analyze the longitudinal plasmonic response of the different geometries. A series of frequency domain field profile monitors were used to examine the electric field patterns in the near field of the nanostructures. The incident light's electric field has a magnitude of 1.0 V/m.

## 3. Results and Discussions

The plasmonic coupling has been studied by changing the nanorod size and the alignment of nanorods (rotational geometries) in the dimer configuration. The study has been carried out to understand and model the coupling between anisotropic nanorods, especially when the aspect ratio (AR) of one nanorod increases by keeping the aspect ratio of another nanorod constant and understanding the effect when nanorods are rotating with each other in the dimer. The longitudinal antibonding mode would not be illuminated for linear homo-dimers, as antiparallel dipoles do not create a net dipole for the dimer [20]. We studied the effects of various factors breaking the structure's homogeneity to explain why antibonding modes appear in the spectrum. The two significant sources of inhomogeneity in dimers are the angular orientation of nanorods (Fig. 1a) and the irregularity of size between the two nanorods (Fig. 1b). The geometry of the dimers is defined by the variable $R = AR_2/AR_1$, where the top and bottom rods are denoted by subscripts 1 and 2.

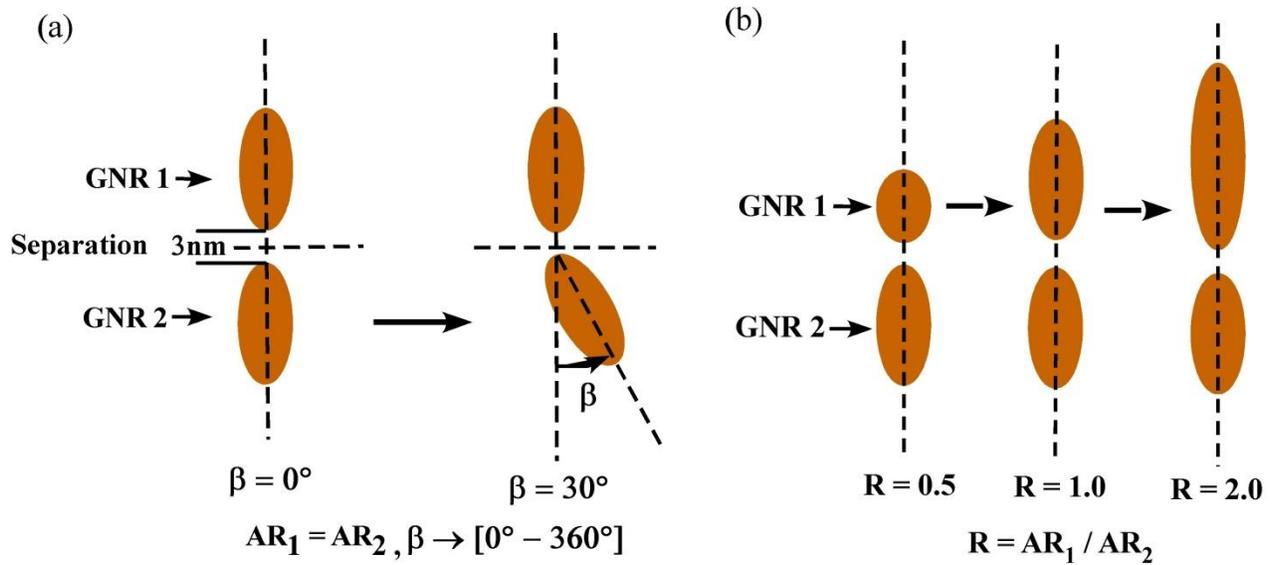

**Fig. 1.** The schematic orientation of nanorods in the dimer (a) with angular rotation, $\beta$ keeping ratio, $R = 1.0$ (b) with two different size nanorods in end-to-end coupling.

We have analyzed the role of the relative orientation of nanorods in dimers on plasmon coupling. To understand the linear geometry deviation in homo-dimers, we have considered nanorods of equal dimensions (70 × 30 nm) with AR = 2.33 and having a separation of 3 nm with an offset angle, $\beta$ between them. We observed the absorption spectra of dimers in Fig. 2, where light polarization is aligned along the nanorod axis, and one nanorod is placed relative to another nanorod at different angular orientations. Bonding and antibonding peaks that arise with angular variation are demonstrated in Fig. 2a and b. As nanorods are in conductive contact, new spectral features for the dimers are evolved, which change the optical properties of dimers. Such conductive contact facilitates plasmon modes which involve the polarization of the charge distribution all over the nanorods and the electric field oscillation between the junctions of nanorods. When a nanorod rotates w.r.t another one by keeping the polarization of the incident light along the nanorod axis, the hybridized plasmon dipole along the nanorod axis becomes blue-shifted due to reduced coupling between nanorods in a dimer. The decoupling of the plasmonic modes happens due to the decrease in the tip-to-tip coupling between the nanorod surfaces as one rod rotates around its center of mass [37].

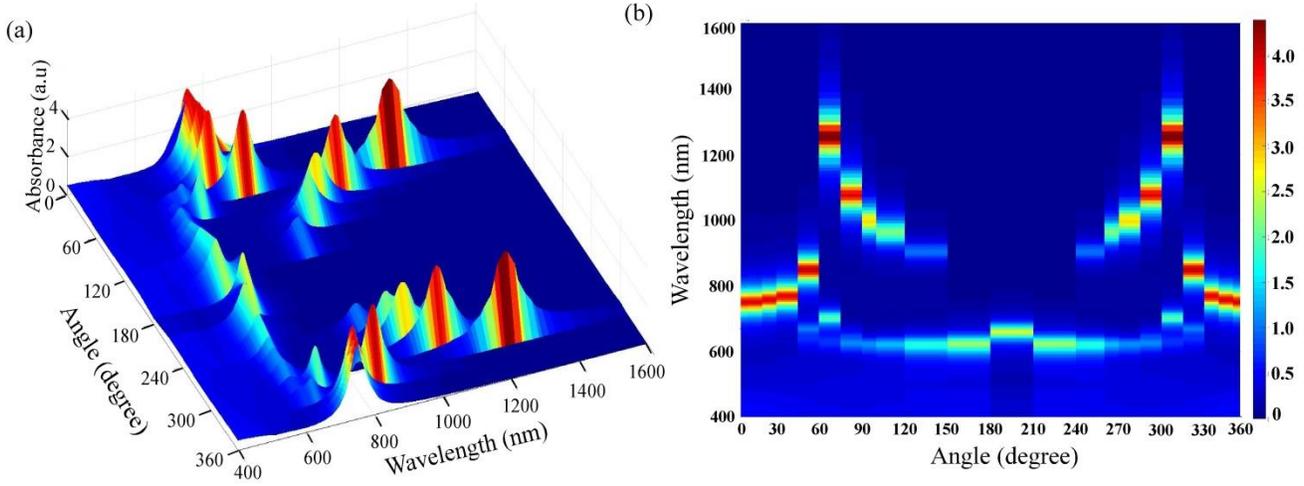

**Fig. 2.** (a) Plasmon coupling vs. angle, β between nanorods in dimer and (b) variation of LSPR peak position w.r.t angular offset, β (top view).

To understand a deep insight into dimers' plasmonic properties, a simple plasmonic hybridization picture is adopted to explain its electromagnetic behavior. The dimers in-plane plasmon modes have been considered in our FDTD calculations. The polarization of light is either parallel or perpendicular to the bisector of angular offset, β inside the substrate plane. In this assumption, four hybridized plasmon modes are expected due to each nanorod's transverse and longitudinal plasmon modes in the dimer [17]. Both modes hybridize separately in alignment with the orientations of end-to-end and side-by-side to provide a bonding and anti-bonding mode when the angle between the nanorods becomes 0° and 180°, respectively. The FDTD calculation shows that the transverse anti-bonding and longitudinal bonding modes are active at 0°, whereas the transverse mode is indistinguishable and shows a single peak around 752 nm. The absorption peaks from the longitudinal anti-bonding and transverse bonding modes are active at 180°, but one single peak at 662 nm appears due to longitudinal polarization. As the size/geometry of the nanorod dimer is reduced, these two modes cannot be described anymore by the combination of individual nanorod modes. Instead, it provides new plasmon modes, which are classified as bonding and anti-bonding modes. Due to the geometry irregularities, the transverse and longitudinal modes are mixed. The observed electric field distributions indicate that the longitudinal dipolar modes of each nanorod ultimately contribute to the hybridized modes [38]. The most exciting feature of bonding and anti-bonding modes of dimers appears in the gap region. With the variation in angular offset, β, as shown in Fig. 3, we have discussed the corresponding variation in LSPR and maximum enhancement in electric fields. The gap region's electrical field is greatly enhanced and reduced when nanorods are resonantly excited about their bonding and anti-bonding modes, respectively.

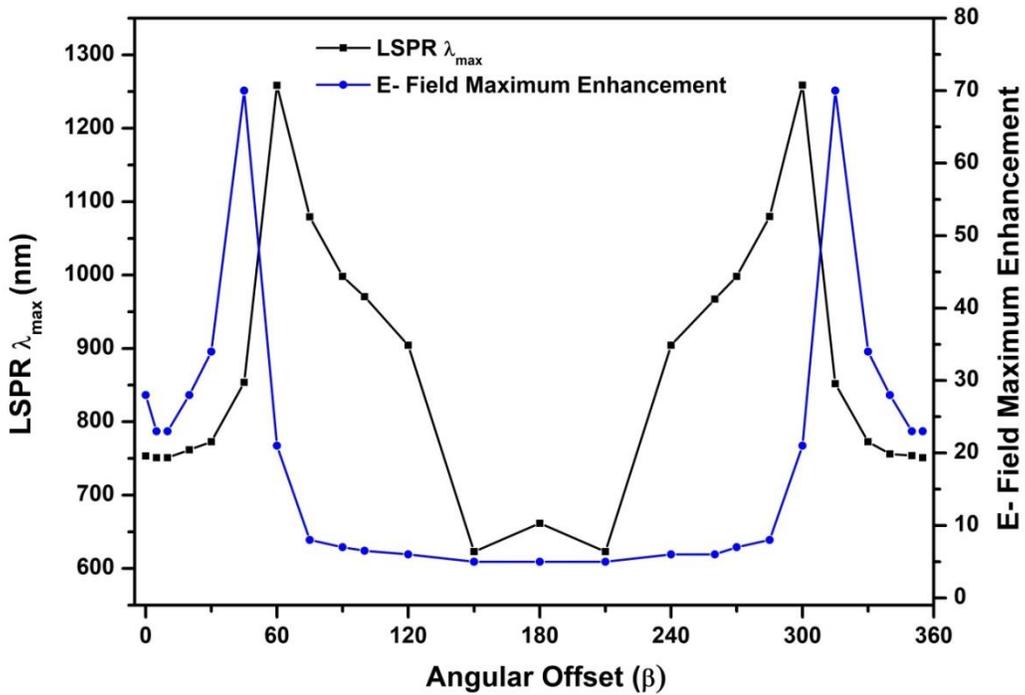

**Fig. 3.** Variation of Longitudinal resonance and electric field maximum enhancement w.r.t. angular offset, β.

Fig. 3 illustrates the symmetry pattern in the properties between the intervals 0˚-180˚ and 180˚-360˚. As a result, we can relate the properties observed in the 0˚-180˚ span to those observed in the 180˚- 360˚. We observed a single absorption peak in the spectrum for the angle, β in the ranges 0˚ < β < 45˚ and 150˚ < β < 180˚ whereas, two absorption peaks have been observed when the angle varies in between ~45˚ and ~150˚. Also, as the angle β increases from 60˚, the strength of the bonding mode decreases, whereas the intensity of the anti-bonding mode increases, which is observed around 625 nm. The bonding mode features are more sensitive as compared to the anti-bonding peak to the angular orientation, β. Only bonding peak is formed for β < 45°, whereas, anti-bonding peak exists for β ≥ 45° in the spectrum. The presence of anti-bonding mode arises from the coupling of the near field of the nanorods and/or the net non-zero dipole moment arising from the tilted dimer structure [39]. The anti-bonding plasmonic mode is characterized by a change in charge that shifts away from the center of the gap of nanorods in dimer configurations. We can correlate the enhancement of the electric field to the bonding mode peak position and intensity. We see an improvement in field strength as the intensity of bonding mode approaches towards angular offset 60°. The intensity of bonding mode decreases after 60°, resulting in reduced field enhancement, and later nearly the same field enhancement in the 120°-240° range, where the antibonding mode is the dominant peak. As the bonding mode dominates again, the rise in electric field intensity displays a maximum increase of around 300°. The electric field enhancement is most significant around 60° and 300°, which is also the highest bonding mode peak angle. The tilted dimer spectrum from 0°< β < 360° clearly shows the different bonding and anti-bonding mode contributions to absorption wavelength, intensity, and electric field enhancement. This symmetry trend between the absorption properties and the angle, β described above could form the basis of nanoscale devices. Devices may be designed to calculate the tilted angles for dimers composed of similar sizes of nanorods.

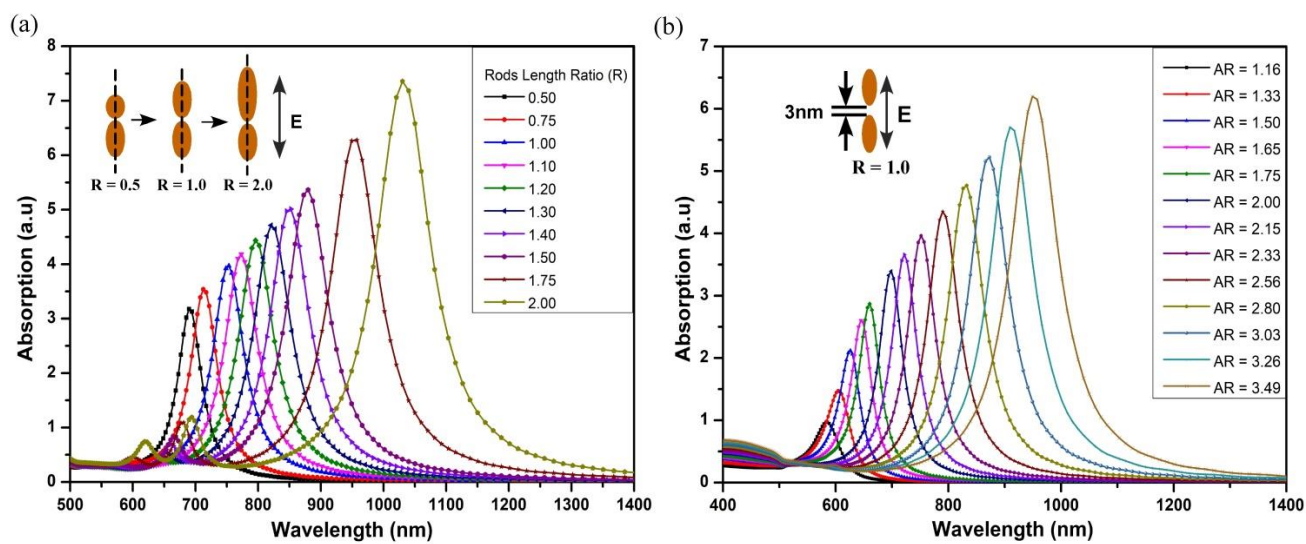

**Fig. 4.** (a) Absorption spectra with increasing R, *i.e.,* increasing the length of one rod and keeping the other rod dimension constant. (b) Variation of LSPR with an increasing aspect ratio of both rods keeping R =1.

Studying the plasmon coupling in the dimers with different dimensions arranged in an end-to-end configuration is interesting. This is important due to the fact that size heterogeneity exists when the dimers are prepared in the laboratory and are of great concern. Our investigation is focused on the size mismatch impact through the FDTD simulation, and the results are shown in Fig. 4a. We observed the sensitivity of the antibonding mode due to its heterogeneity in size. The geometry of the dimers is described by the ratio, R, which is the ratio of the aspect ratio of the nanorods in the dimer. It has been found that the aspect ratio of the nanorod is an essential parameter of the longitudinal plasmon energy since the LSPR is approximately a linear function of it. The previously laboratory synthesized nanorod sample have the same average diameter and varying average lengths. So in our simulations, gold nanorods having the same diameter but different lengths have been considered. To change the ratio, R, the length of the bottom nanorod was kept fixed while the length of the top nanorod was altered. As an increase in dimer size, the electromagnetic field impacted on dimer can no longer be believed to have a small phase difference across the NRs length. Hence, the conduction electrons cannot oscillate collectively. The phase retardation effect occurs when the conduction electrons on the facing and opposing sides of the nanorod oscillate with different phases, resulting in the stimulation of a new plasmon mode. Based on the AR of the nanorod, Au supports multiple plasmon modes throughout the spectral range. A close look at the spectral distribution (between 640 and 700 nm) in Fig. 4a shows very weak bumps originating from the longitudinal anti-bonding mode. The spectrum displays only a single peak corresponding to the bonding mode for R up to 1.1. The antibonding dimer mode became prominent for heterodimers with R=1.2 and was placed on the blue side of the bonding mode. As the R-value increases (R $\geq$ 1.2), the bonding mode redshifts, the antibonding mode starts to emerge, and redshifts with higher intensity. A significant absorption peak (see Fig. 4b), which arises from the plasmon modes of longitudinal bonding, is observed above 700 nm for all the homo-dimers (R = 1). When the rods are connected, the opposite charges are firmly localized at the ends of the two rods at the gap. As a result, strong attraction across the gap competes with the intrarod restoring forces that drive intrarod charge oscillations, resulting in a red shift in LSPR. For dimers with R $\neq$ 1, the mirror geometry is no more holds good, leading to a non-zero dipole moment for the antibonding modes, and as a result, a weak peak is observed on the higher energy side of the absorption spectrum. In any dimer, if the absorption cross-section of one nanoparticle is slightly greater than that of the neighboring nanoparticle, then the absorption properties would be dominated by the larger nanoparticle irrespective of which nanoparticle gets excited first [26]. The coupled plasmon resonance is red-shifted compared to the plasmon resonance of individual nanorods. This finding shows that even if the distance between the nanorods is kept constant, the coupled plasmon energy of the dimer can be regulated desirably by adjusting the size of the individual metal nanorods. It has been understood that the plasmonic interactions are highly dependent on the configuration of nanorods in the dimer. When excited by polarized light along the length of the nanorods in end-to-end homo-dimers, it is red-shifted coupled plasmon mode whereas, for side-by-side homo-dimers, it is significantly blue-shifted. For the linear, end-to-end structure, the coupled plasmon mode has a bonding nature, and it has an anti-bonding character for the side-by-side configuration.

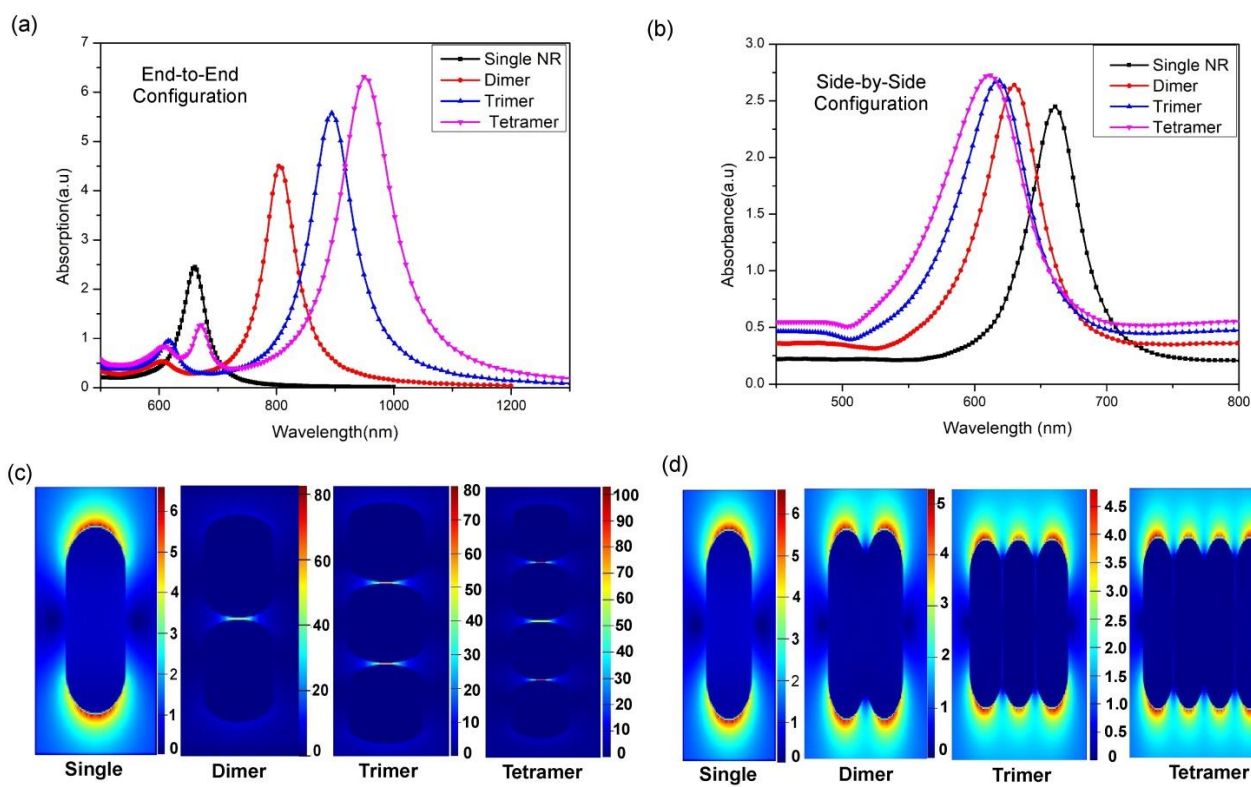

**Fig. 5**. Absorption spectrum of different nanostructure geometry of aspect ratio 2.33 with (a) end-to-end (b) side-by-side configurations. Electric field mapping of the orientation in (c) end-to-end (d) side-by-side configurations.

A few simulations for various configurations were done to acquire an idea of different dependence on influencing variables to understand better how the improvements are connected in the involvement of multiple GNRs. Also, while not all of the configurations are experimentally achievable, it provides some insight into how they could give rise to a coupled field. It would provide an overview of the various assemblies needed to show other interactions in a given system. Now a comparative analysis of single NR, dimer, trimer, and tetramer orientation of identical nanorods of aspect ratio 2.33 has been done to understand a larger structure with more number of NRs. We analyzed the side-by-side and end-to-end configuration structures. The gold nanorods are positioned end-to-end in various forms with a separation of 1 nm. The single nanorod shows longitudinal plasmon resonance around ~ 660 nm (Fig. 5a). About ~ 605 nm and ~ 805 nm, two plasmon peaks are provided by the dimer configuration. Likewise, the trimer shape also gives two extremes around ~ 615 nm and ~ 895 nm, and there are peaks at 608 nm, 672 nm, and 950 nm in the tetramer form. The second peak around 672 nm is due to more hot spots created by the assembly structures. As a result, more particles forming the same configuration in a single chain will strengthen the plasmon resonance to a higher wavelength. We can see that the plasmon resonance increases from 660 nm to 950 nm as the arrangement differs from single particle to tetramer type. When GNRs are adjacent, the electric field enhancement is often significant; this enables the local field to be coupled and contributes to higher field values in the area between the GNRs. As shown in Fig. 5c, the electric field increased in the dimer configuration compared to the single nanorod with a maximum enhancement factor of 6. Due to the electric field, the electric charge oscillates in the chain, and the maximum field intensity is always distributed in the gap zone. Hence, as NRs get closer to each other, the opposite charges distribute alternatively, leading to increased interactions. It has been found that the electric field is enhanced many folds in a specific point, called hot spots, compared to the rest of the system.

The absorption spectra of the gold nanorods (with AR= 2.33), configured in a side-by-side orientation with a separation of 1 nm, are shown in Fig. 5b. The dimer configuration gives a 630 nm plasmon peak, whereas the side-by-side trimer of GNRs shows plasmon resonance around ~ 615 nm, while the tetramer of GNRs shows a peak around ~ 612 nm. These side-by-side assembly configurations are blue-shifted in the plasmon resonance from 660 nm to 612 nm. The electric field enhancement is weaker when GNRs are closer in a side-by-side ensemble, as shown in Fig. 5d. In the higher configuration, we observe the diminished electric field related to the single nanorod and equivalent for all higher configurations. Therefore, the electric field enhancement is not influenced by this side-by-side assembly configuration. The highest field strength is at the edge surface of the nanorod in the side-by-side assembly.

We investigated the coupling strength between the longitudinal plasmon resonances in various end-to-end directed structures containing gold nanorods (n = 4 - 16). We observed the overall influence of the various nanostructures on the plasmon coupling and compared it with the dimer, trimer, and tetramer assembly structure. The spacing between the nanorods has been kept at 1 nm so that the near-fields of the individual nanorods interact strongly. We display the respective absorption spectra for different structures in Fig. 6a and their electric field distribution in Fig. 6b. As shown in Fig. 6a, the LSPR redshifts with an increasing number of nanorods with the same orientation type, as it varies from other structures like 5 NRs. Therefore, plasmon coupling relies on the orientation of the particle for the same size and shape. We observed that increasing the number of particles increases the optical strength, reducing their multiple nanorods structure. The multiple side-by-side trimer structure with the same orientation type, i.e. (3, 6, 9) with an odd number of nanorods, indicates that the peak value is in the same range with a variation of ± 4 nm, and the peak intensity decreases when the structure becomes larger. Similarly, with the same type of orientation, i.e. (4, 8, 12, 16) with an even number of nanorods, the multiples of the side-by-side tetramer structure indicate that the peak value is in the same range with a variation of ± 5 nm, while the peak intensity decreases when the system becomes larger. Therefore, the large cluster of nanorods can be related to their first nanorods of the same direction as they exhibit almost similar optical absorption and electric field enhancement.

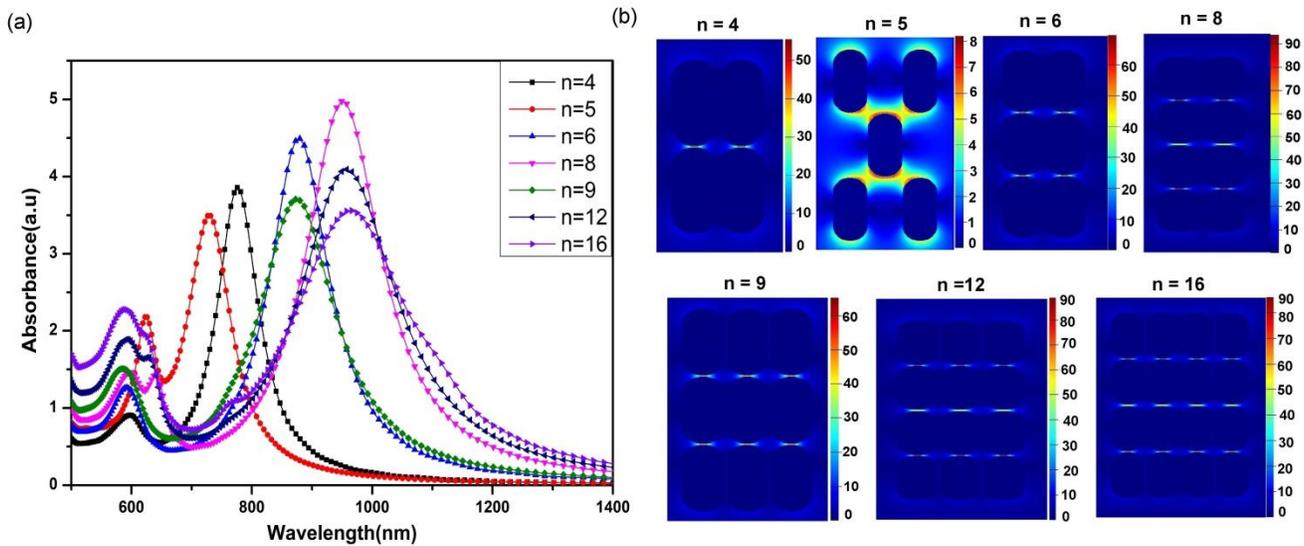

**Fig. 6.** (a) Absorption spectra and (b) electric field distribution mapping of different nanostructures with 'n' number of gold nanorod in various arrays.

The optical properties of a 4 × 4 periodic matrix structure of gold nanorods separated by a distance of 1 to 70 nm have been investigated, and the results are shown in Fig. 7. We found that the absorption cross-section does not exhibit resonances with consistent symmetrical line profiles. We tried to understand the properties based on particle spacing, which correlates with the electric field enhancements. We increased the separation among the nanorods from 1 to 70 nm to understand the coupling between the light and resonance modes leading to Fano-like resonance. In turn, the array of nanorods exhibits some hybrid modes arising from plasmon resonances of the various nanorods. The hybrid modes resonate as bonding mode at a higher wavelength and as an anti-bonding mode at a lower wavelength with an FWHM much broader for the bonding mode than the anti-bonding mode (see Fig. 7a). These hybrid modes resonate at 965 nm and 585 nm, with HWHMs of 208 nm and 29 nm at a spacing of 1 nm, respectively. The bonding mode blue-shifted exponentially from 965 nm to 670 nm (Fig. 7b) and the anti-bonding mode red-shifted from 585 nm to 650 nm with increasing resonance peak intensity. The electric field enhancement of the system also decreases exponentially by increasing the gap from 1 to 70 nm, as shown in Fig. 7b. We can see from Fig. 7c that the interaction between the rods decreases with increasing separation distance, and the individual features become prominent with no significant interaction among the individual rods. The separation between charges on opposed surfaces increases as particle gap widths increase. The Coulomb interaction between them is attenuated, resulting in a shorter Surface plasmon resonance wavelength and a blue shift in the spectra. We discovered that the structure's spectral pattern and Fano resonance are susceptible to the spacing.

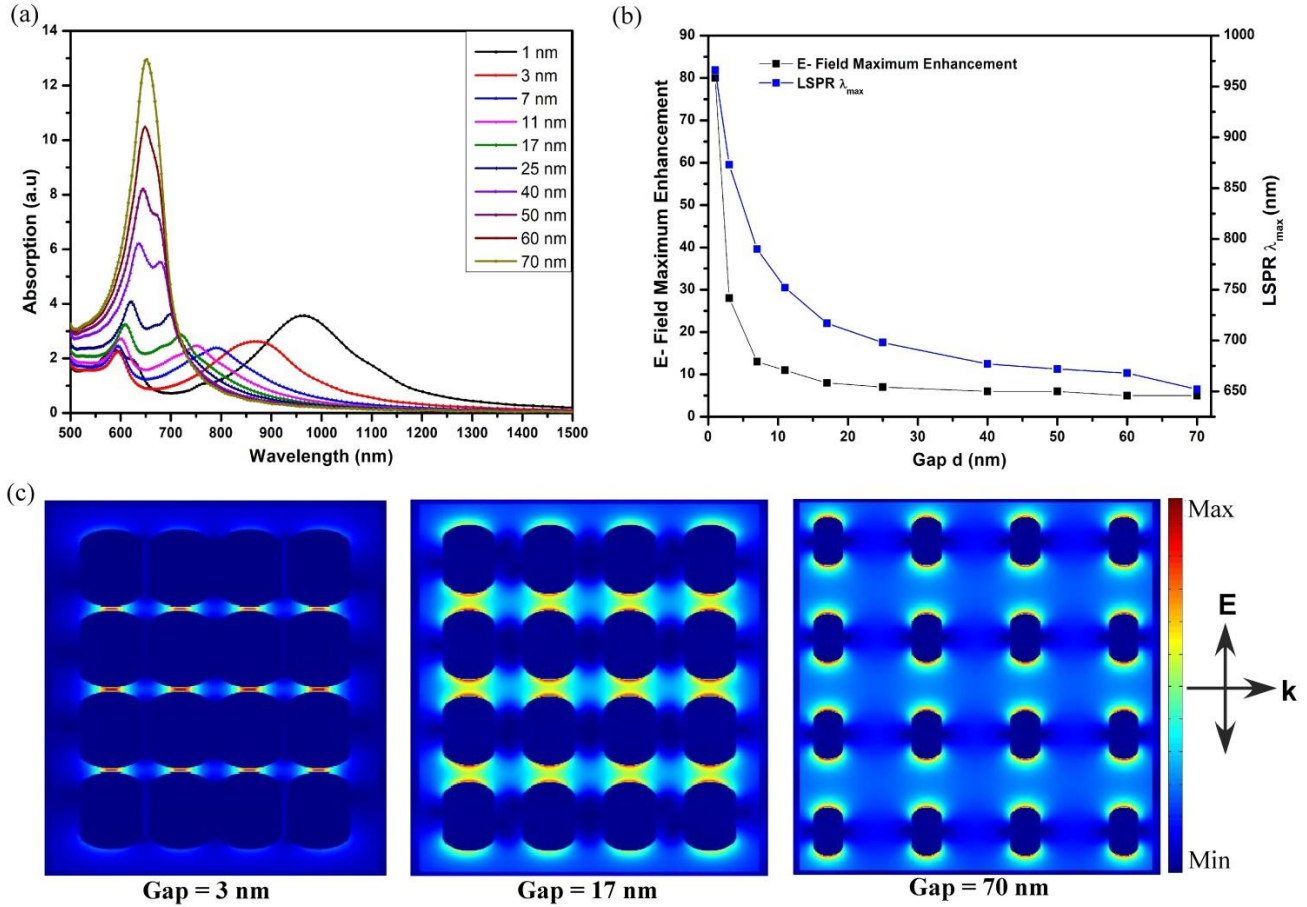

**Fig. 7.** (a) Absorption spectra of 4 × 4 array structure with the variation of the gap, (b) variation of LSPR and electric field enhancement with the separation of the gap, *d* in nanoparticles, and (c) electric field mapping at different separation gap *namely* 3 nm, 17 nm, and 70 nm.

The resonant modes' spectral overlap can explain the fano-like resonance, and the asymmetry primarily arises from the bonding coupling phase of the above array structure. The configurations size of nanorods in the system and the conditions of illumination influence both the absorption and coupling of individual resonant modes. The resultant coupling between plasmon resonances of NRs may cause the group of particles to undergo mutual coupling, contributing to hybrid modes. This is well reflected in the total optical properties of the array structures [40]. The surface charge density distributions on the NRs and the derived electric moments determine the different mode resonances. The Fano dip of the absorption spectrum slowly disappears as the distance between the nanorods increases, and the spectrum transforms into a single Lorentzian resonance band. The electric field enhancement contributions often indicate a shift from asymmetric to symmetric line profiles as the separation increases in the structure.

The result suggests that a minimum gap size is needed to achieve a Fano-type resonance. It is possible to view the variation in the resonance line profiles by increasing the nanorods gap. In the separation gap of ≥ 60 nm, the coupling between these nanorods is no longer strong enough to distinguish both modes and exhibit a Lorentzian characteristic of an independent nanorod. The increase in the separation distance among nanorods results in blue shifts followed by exponential decay behavior, shortening the fano dip in the absorption spectra. The contribution of various modes to the absorption spectra indicates an increase in the relative intensity of the peak with a decrease in the electric field by increasing the separation distance. As demonstrated prior that a particular structure size is necessary for forming a Fano-like resonance, we investigate how such optical characteristics of the structure change as its dimension increases. When the size of the nanorod is kept constant while the spacing gap between particles is evenly increased, the extinction spectra undergo a blue shift, and the Fano dip narrows. In general, nanoparticle size, spatial configuration, and illumination settings influence both direct excitation and mutual coupling between resonant modes, determining the plasmonic structure's overall spectral response. These behaviors illustrate the importance of separation between nanorods in array size to optical efficiency, which is critical in structure design for relevant applications.

## 4. Conclusion

The plasmon coupling between gold nanorods depends on their relative orientation and interparticle spacing. A detailed understanding of plasmon coupling in gold nanorod dimers in different geometry has been studied using the FDTD simulation methodology. We discussed the response by varying the size heterogeneity and rotational geometries. Plasmon hybridization corresponds to anti-bonding, and bonding modes have been observed in our studies. The plasmon coupling has also been studied by varying the aspect ratio of one nanorod and keeping another nanorod constant with a small spacing between the nanorods. The increase in spacing leads to a decrease in the electric field enhancement. In a heterodimer, anti-bonding dimer modes that would not be noticeable compared to the homodimers emerge clearly in the spectrum. The angular-related coupling shows that the electric field enhancement happens only when the bonding plasmon mode is excited by external light. The large nanorod ensembles may be compared to their small ensemble structure of the same orientation, as the optical absorption and electric field enhancement are precisely in the same region. This plasmonic coupling of different sizes (monomers, dimers, trimers, tetramers) of nanorods presents a better tool for tuning the frequency, intensity, spatial distribution, and polarization of local electric field inside and around nanostructures.

There is an interaction between broad and narrow resonant modes in gold nanorods' 4 × 4 array structure, resulting in a Fano resonance with its distinctively asymmetric line shape. The extinction spectrum's signature Fano dip gradually disappears as the distance between the nanorods increases and turns into a single Lorentzian-shaped resonance band. We observed that plasmon coupling depends on four factors: the distance of separation, orientation, size heterogeneity, and the number of nanorods in the geometry. Understanding the dependence on the number, distance, and orientation between the nanorods of collective plasmonic properties provides a functional layout for developing different applications.

**Funding**

The author, A. K. Sahu, receive financial support from CSIR, the Government of India.

**CRediT authorship contribution statement**

**Aditya K. Sahu**: Conceptualization, Methodology, Software, Data analysis, Investigation, Writing - original draft, Writing, Writing - reviewing & editing. **Satyabrata Raj**: Resources, Writing - reviewing & editing, Funding acquisition, Supervision.

**Declaration of competing interests**

The authors declare that they have no known competing financial interests or personal relationships that could have influenced the work reported in this paper.

**Availability of Data and Material**

The data that support the findings of this study are available from the corresponding author upon reasonable request.

**References**

[1] S. Nie, S. R. Emory, Probing Single Molecules and Single Nanoparticles by Surface-Enhanced Raman Scattering, Science 275 (1997) 1102-1106, https://doi.org/10.1126/science.275.5303.1102.

[2] X. Wang, L. Yao, X. Chen, H. Dai, M. Wang, L. Zhang, Y. Ni, L. Xiao, J. Han, Gap-Induced Giant Third-Order Optical Nonlinearity and Long Electron Relaxation Time in Random-Distributed Gold Nanorod Arrays, ACS Appl. Mater. Interfaces 11 (2019) 32469-32474, https://doi.org/10.1021/acsami.9b08935 .

[3] V. Krivenkov, S. Goncharov, P. Samokhvalov, A. Sanchez-Iglesias, M. Grzelczak, I. Nabiew, Y. Rakovich, Enhancement of Biexciton Emission Due to Long-Range Interaction of Single Quantum Dots and Gold Nanorods in a Thin-Film Hybrid Nanostructure, J. Phys. Chem. Lett. 10 (2019) 481−486, https://doi.org/10.1021/acs.jpclett.8b03549.

[4] Y. J. Bao, Y. Yu, H. F. Xu, Q. L . Lin, Y. Wang, J. T. Li, Z. K. Zhou, X. Wang, Coherent Pixel Design of Metasurfaces for Multidimensional Optical Control of Multiple Printing-Image Switching and Encoding, Adv. Funct. Mater. 28 (2018) 1805306, https://doi.org/10.1002/adfm.201805306 .

[5] L. H. Yao, J. P. Zhang, H. W. Dai, M. S. Wang, L. M. Zhang, X. Wang, J. B. Han, Plasmon-enhanced versatile optical nonlinearities in an Au–Ag–Au multi-segmental hybrid structure, Nanoscale 10 (2018) 12695−12703, https://doi.org/10.1039/C8NR02938E.

[6] H. Dai, L. Zhang, Z. Wang, X. Wang, J. Zhang, H. Gong, J. B. Han, Y. Han, Linear and Nonlinear Optical Properties of Silver-Coated Gold Nanorods, J. Phys. Chem. C 121 (2017) 12358−12364, https://doi.org/10.1021/acs.jpcc.7b00295 .

[7] Z. W. Ma, C. Chi, Y. Yu, Z. Q. Zhong, L. H. Yao, Z. K. Zhou, X. Wang, Y. B. Han, J. B. Han, Near-UV-enhanced broad-band large third-order optical nonlinearity in aluminum nanorod array film with sub-10 nm gaps, Opt. Express 24 (2016) 5387−5394, https://doi.org/10.1364/OE.24.005387 .

[8] Z. K. Zhou, J. Liu, Y. J. Bao, L. Wu, C. E. Png, X. H. Wang, C. W. Qiu, Quantum plasmonics get applied, Prog. Quantum. Electron. 65 (2019) 1−20, https://doi.org/10.1016/j.pquantelec.2019.04.002

[9] L. Zhang, H. Dai, X. Wang, L. Yao, Z. Ma, J. B. Han, Nonlinear optical properties of Au–Ag core–shell nanorods for all-optical switching, J. Phys. D: Appl. Phys. 50 (2017) 355302, https://doi.org/10.1088/1361-6463/aa7c51 .

[10] J. Li, S. K. Cushing, F. Meng, T. R. Senty, A. D. Bristow, N. Wu, Plasmon-induced resonance energy transfer for solar energy conversion, Nat. Photonics 9 (2015) 601−607, https://doi.org/10.1038/nphoton.2015.142 .

[11] M. T. Sheldon, J. van de Groep, A. M. Brown, A. Polman, H. A. Atwater, Plasmoelectric potentials in metal nanostructures, Science 346 (2014) 828−831, https://doi.org/10.1126/science.1258405.

[12] U. Kreibig, M. Vollmer, Optical Properties of Metal Clusters; Springer: Berlin, 1995.

[13] K. L. Kelly, E. Coronado, L. L. Zhao, G. C. Schatz, The Optical Properties of Metal Nanoparticles: The Influence of Size, Shape, and Dielectric Environment, J. Phys. Chem. B 107 (2003) 668–677, https://doi.org/10.1021/jp026731y .

[14] S. Sheikholeslami, Y. W. Jun, P. K. Jain, A. P. Alivisatos, Coupling of Optical Resonances in a Compositionally Asymmetric Plasmonic Nanoparticle Dimer, Nano Lett. 10 (2010) 2655-2660, https://doi.org/10.1021/nl101380f.

[15] L. Shao, Q. Ruan, R. Jiang, J. Wang, Macroscale Colloidal Noble Metal Nanocrystal Arrays and Their Refractive Index-Based Sensing Characteristics, Small 10 (2014) 802−811, https://doi.org/10.1002/smll.201301812.

[16] J. Lee, M. Tymchenko, C. Argyropoulos, P. Y. Chen, F. Lu, F. Demmerle, G. Boehm, M. C. Amann, A. Alu, M. A. Belkin, Giant nonlinear response from plasmonic metasurfaces coupled to intersubband transitions, Nature 511 (2014) 65−69, https://doi.org/10.1038/nature13455.

[17] L. Shao, K. C. Woo, H. Chen, Z. Jin, J. Wang, H. Q. Lin, Angle- and Energy-Resolved Plasmon Coupling in Gold Nanorod Dimers, ACS Nano 4 (2010) 3053−3062, https://doi.org/10.1021/nn100180d.

[18] R. deWaele, A. F. Koenderink, A. Polman, Tunable Nanoscale Localization of Energy on Plasmon Particle Arrays, Nano Lett. 7 (2007) 2004–2008, https://doi.org/10.1021/nl070807q.

[19] J. Zuloaga, E. Prodan, P. Nordlander, Quantum Description of the Plasmon Resonances of a Nanoparticle Dimer, Nano Lett. 9 (2009) 887–891, https://doi.org/10.1021/nl803811g.

[20] L. S. Slaughter, Y. Wu, B. A. Willingham, P. Nordlander, S. Link, Effects of Symmetry Breaking and Conductive Contact on the Plasmon Coupling in Gold Nanorod Dimers, ACS Nano 4 (2010) 4657-4666, https://doi.org/10.1021/nn1011144.

[21] E. Prodan, C. Radloff, N. J. Halas, P. Nordlander, A Hybridization Model for the Plasmon Response of Complex Nanostructures, Science 302 (2003) 419−422, https://doi.org/10.1126/science.1089171.

[22] J. H. Yoon, F. Selbach, L. Schumacher, J. Jose, S. Schlücker, Surface Plasmon Coupling in Dimers of Gold Nanoparticles: Experiment and Theory for Ideal (Spherical) and Nonideal (Faceted) Building Blocks, ACS Photonics 6 (2019) 642-648, https://doi.org/10.1021/acsphotonics.8b01424

[23] W. Li, P. H. C. Camargo, X. Lu, Y. Xia, Dimers of Silver Nanospheres: Facile Synthesis and Their Use as Hot Spots for Surface-Enhanced Raman Scattering, Nano Lett. 9 (2009) 485–490, https://doi.org/10.1021/nl803621x.

[24] S. Kim, J. Jin, Y. J. Kim, I. Y. Park, Y. Kim, S. W. Kim, High-harmonic generation by resonant plasmon field enhancement, Nature 453 (2008) 757–760, https://doi.org/10.1038/nature07012.

[25] A. Kinkhabwala, Z. Yu, S. Fan, Y. Avlasevich, K. Mullen, W. E. Moerner, Large single-molecule fluorescence enhancements produced by a bowtie nanoantenna, Nat. Photonics 3 (2009) 654–657, https://doi.org/10.1038/nphoton.2009.187 .

[26] K. Ueno, S. Juodkazis, V. Mizeikis, K. Sasaki, H. Misawa, Clusters of Closely Spaced Gold Nanoparticles as a Sourceof Two-Photon Photoluminescence at Visible Wavelengths, Adv. Mater. 20 (2008) 26–30, https://doi.org/10.1002/adma.200602680 .

[27] A. N. Grigorenko, N. W. Roberts, M. R. Dickinson, Y. Zhang, Nanometric optical tweezers based on nanostructured substrates, Nat. Photonics 2 (2008) 365–370, https://doi.org/10.1038/nphoton.2008.78

[28] P. Muhlschlegel, H. J. Eisler, O. J. F. Martin, B. Hecht, D. W. Pohl, Resonant Optical Antennas, Science 308 (2005) 1607–1609, https://doi.org/10.1126/science.1111886 .

[29] M. Schnell, A. Garcia-Etxarri, A. J. Huber, K. Crozier, J. Aizpurua, R. Hillenbrand, Controlling the near-field oscillations of loaded plasmonic nanoantennas, Nat. Photonics 3 (2009) 287–291, https://doi.org/10.1038/nphoton.2009.46 .

[30] P. K. Jain, S. Eustis, M. A. El-Sayed, Plasmon Coupling in Nanorod Assemblies: Optical Absorption, Discrete Dipole Approximation Simulation, and Exciton-Coupling Model, J. Phys. Chem. B 110 (2006) 18243–18253, https://doi.org/10.1021/jp063879z .

[31] M. Lisunova, J. Norman, P. Blake, G. T. Forcherio, D. F. DeJarnette, D. K. Roper, Modulation of plasmonic Fano resonance by the shape of the nanoparticles in ordered arrays, J. Phys. D: Appl. Phys. 46 (2013) 485103, https://doi.org/10.1088/0022-3727/46/48/485103 .

[32] M. R. Shcherbakov, P. P. Vabishchevich, V. V. Komarova, T. V. Dolgova, V. I. Panov, V. V. Moshchalkov, A. A. Fedyanin, Ultrafast Polarization Shaping with Fano Plasmonic Crystals, Phys. Rev. Lett. 108 (2012) 253903, https://doi.org/10.1103/PhysRevLett.108.253903 .

[33] S. Zou, G. C. Schatz, Silver nanoparticle array structures that produce giant enhancements in electromagnetic fields, Chem. Phys. Lett. 403 (2005) 62-67, https://doi.org/10.1016/j.cplett.2004.12.107

[34] Z. L. Samson, K. F. Macdonald, F. D. Angelis, B. Gholipour, K. Knight, C. C. Huang, E. D. Fabrizio, D. W. Hewak, N. I. Zheludev, Metamaterial electro-optic switch of nanoscale thickness, Appl. Phys. Lett. 96 (2010) 143105, https://doi.org/10.1063/1.3355544 .

[35] Lumerical Solutions, Inc. (trial version), www.docs.lumerical.com/en/fdtd/reference_guide.html .

[36] P. B. Johnson, R. W. Christy, Optical Constants of the Noble Metals, Phys. Rev. B 6 (1972) 4370 – 4379, https://doi.org/10.1103/PhysRevB.6.4370 .

[37] C. Tabor, D. V. Haute, M. A. El-Sayed, Effect of Orientation on Plasmonic Coupling between Gold Nanorods, ACS Nano 3 (2009) 3670-3678, https://doi.org/10.1021/nn900779f .

[38] J. Wu, X. Lu, Q. Zhu, J. Zhao, Q. Shen, L. Zhan, W. Ni, Angle-Resolved Plasmonic Properties of Single Gold Nanorod Dimers, Nanomicro Lett. 6 (2014) 372–380, https://doi.org/10.1007/s40820-014-0011-7 .

[39] S. Biswas, D. Nepal, K. Park, R. A. Vaia, Orientation Sensing with Color Using Plasmonic Gold Nanorods and Assemblies, J. Phys. Chem. Lett. 3 (2012) 2568−2574, https://doi.org/10.1021/jz3009908 .

[40] S. Bakhti, A. V. Tishchenko, X. Zambrana-Puyalto, N. Bonod, S. D. Dhuey, P. J. Schuck, S. Cabrini, S. Alayoglu, N. Destouches, Fano-like resonance emerging from magnetic and electric plasmon mode coupling in small arrays of gold particles, Sci. Rep. 6 (2016) 32061, https://doi.org/10.1038/srep32061 .